\documentclass[twocolumn]{aastex62}

\usepackage{natbib}

\submitjournal{ApJ}

\shorttitle{The Fifth IMBH Candidate in the Galactic Center}
\shortauthors{Takekawa et al.}
\accepted{2020 January 22}

\begin{document}

\title{The Fifth Candidate for an Intermediate-mass Black Hole in the Galactic Center}

\correspondingauthor{Shunya Takekawa}
\email{shunya.takekawa@nao.ac.jp}

\author{Shunya Takekawa}
\affil{Nobeyama Radio Observatory, National Astronomical Observatory of Japan\\
462-2 Nobeyama, Minamimaki, Minamisaku-gun, Nagano 384-1305, Japan}

\author{Tomoharu Oka}
\affiliation{Department of Physics, Institute of Science and Technology, Keio University\\
3-14-1 Hiyoshi, Kohoku-ku, Yokohama, Kanagawa 223-8522, Japan}
\affiliation{School of Fundamental Science and Technology, Graduate School of Science and Technology, Keio University\\
3-14-1 Hiyoshi, Kohoku-ku, Yokohama, Kanagawa 223-8522, Japan}

\author{Yuhei Iwata}
\affiliation{School of Fundamental Science and Technology, Graduate School of Science and Technology, Keio University\\
3-14-1 Hiyoshi, Kohoku-ku, Yokohama, Kanagawa 223-8522, Japan}

\author{Shiho Tsujimoto}
\affiliation{School of Fundamental Science and Technology, Graduate School of Science and Technology, Keio University\\
3-14-1 Hiyoshi, Kohoku-ku, Yokohama, Kanagawa 223-8522, Japan}

\author{Mariko Nomura}
\affiliation{Faculty of Natural Sciences, National Institute of Technology, Kure College\\
2-2-11 Agaminami, Kure, Hiroshima 737-8506, Japan}

\begin{abstract}
We report the results of high-resolution molecular line observations of the high-velocity compact cloud HCN--0.085--0.094 with the Atacama Large Millimeter/submillimeter Array.
The HCN {\it J}=4--3, HCO$^+$ {\it J}=4--3, and CS {\it J}=7--6 line images reveal that HCN--0.085--0.094 consists mainly of three small clumps with extremely broad velocity widths.
Each of the three clumps has a 5.5 GHz radio continuum counterpart in its periphery toward Sgr A$^*$.
The positional relationship indicates that their surfaces have been ionized by ultraviolet photons from young stars in the central cluster, suggesting the clumps are in close proximity to the Galactic nucleus.
One of the three clumps has a ring-like structure with a very steep velocity gradient.
This kinematical structure suggests an orbit around a point-like object with a mass of $\sim 10^4$ $M_\odot$.
The absence of stellar counterparts indicates that the point-like object may be a quiescent black hole.
This discovery adds another intermediate-mass black hole candidate in the central region of our Galaxy.

\end{abstract}


\section{Introduction}
High-velocity compact clouds (HVCCs) are a peculiar population of compact ($d\lesssim 5$ pc) molecular clouds with extremely broad velocity widths ($\Delta V \gtrsim 50$ km s$^{-1}$) in the Central Molecular Zone (CMZ) of our Galaxy \citep{oka98, oka12, tokuyama19}.
More than 80 HVCCs have been identified in the CMZ based on the CO {\it J}=1--0 observations with the Nobeyama Radio Observatory 45 m telescope \citep{nagai08}.
Most of the HVCCs have no counterparts at other wavelengths, rendering it difficult to understand their origins.
Interactions with supernovae \citep{oka99, oka08, tanaka09, yalinewich18} and cloud--cloud collisions \citep{tanaka15, tanaka18, ravi18} are possible explanations for HVCCs, while gravitational interactions between molecular clouds and massive ($> 10^4$ $M_\odot$) point-like objects have been suggested for some HVCCs \citep{oka16, oka17, ballone18, takekawa19a, takekawa19b}.
These massive point-like objects are candidates for intermediate-mass black holes (IMBHs).

 Although there is no definitive evidence for the existence of IMBHs yet, a number of IMBH candidates have been suggested so far \citep{mezcua17, koliopanos17}.
Ultraluminous X-ray sources have been proposed as IMBHs with high accretion rates \citep[e.g.,][]{farrell09}.
Several dwarf galaxies have been suggested to harbor active galactic nuclei with massive ($10^4$--$10^5$~$M_\odot$) IMBHs \citep[e.g.,][]{filippenko03, reines13}.
Many more dwarf galaxies may hold IMBHs with low accretion rates, which are hardly detectable by electromagnetic observations \citep{bellovary19}.
Dynamical analyses of globular clusters have also suggested that IMBHs possibly lurk at their centers \citep[e.g.,][]{gebhardt02, kiziltan17}.
Numerical simulations have shown that IMBHs could be formed through runaway collisions of stars in dense star clusters \citep[e.g.,][]{portegieszwart02}.
Such IMBHs would be brought into the Galactic center with their parent clusters through dynamical friction, and may possibly be moving around the Galactic nucleus \citep{fujii09, arca-sedda18, arca-sedda19}.

HVCCs can be potential probes of such wandering IMBHs \citep[e.g.,][]{takekawa19a}.
This is evidenced by our high-resolution observations of the HVCC HCN--0.044--0.009 \citep{takekawa17b} with the Atacama Large Millimeter/submillimeter Array (ALMA), which revealed that HCN--0.044--0.009 actually consists of two dense molecular gas streams in Keplerian motion around an invisible mass of $(3.2\pm0.6)\times 10^4$ $M_\odot$.
This is a promising candidate for an IMBH \citep{takekawa19a}.
CO--0.40--0.22 \citep{oka16}, CO--0.31+0.11 \citep{takekawa19b}, and IRS13E \citep{tsuboi17, tsuboi19}, all in the Galactic center, have also been proposed to harbor massive IMBHs based on their broad-velocity-width nature.

Another HVCC known as HCN--0.085--0.094 has originally been suggested to be driven by a high-velocity plunge of a black hole into a molecular cloud \citep{takekawa17b, nomura18}. HCN--0.085--0.094 is at a projected distance of $\sim 8$ pc from the Galactic nucleus Sgr A$^*$ \citep{takekawa17b}. This paper reports on the results of the ALMA high-resolution observations of HCN--0.085--0.094 and the discovery of a very compact molecular clump showing a signature of a rotational motion.
The distance to the Galactic center is assumed to be $D=8$ kpc.

\section{Observations}
Our ALMA cycle 5 observations of HCN--0.085--0.094 (2017.1.01557.S) were performed in 2018 May with eleven 7 m antennas (on May 13), forty-six 12 m antennas (on May 14), and four Total Power (TP) antennas (on May 24 and 28).
HCN--0.009--0.044 was also observed in this project \citep{takekawa19a}.
The field of view for HCN--0.085--0.094 was $52\arcsec \times 62\arcsec$ centered at $(l,\ b)=(-0\arcdeg.083,\ -0\arcdeg.091)$, which was covered with 11 and 27 pointings of the 7 m and 12 m arrays, respectively.
The 12 m array observations were operated in the C43-2 configuration with baseline lengths of 15--314 m.
The target lines were HCN {\it J}=4--3 (354.5 GHz), HCO$^+$ {\it J}=4--3 (353.6 GHz), and CS {\it J}=7--6 (342.9 GHz).
The bandwidths of the spectral windows were 1, 0.5, and 2 GHz with a 1.953 MHz channel width for the HCN, HCO$^+$, and CS observations, respectively.
The on-source times were 51.74, 10.89, and 186.48 min for the 7 m, 12 m, and TP array observations, respectively.
J1924--2914 and J1517--2422 were used for the flux and bandpass calibrations.
The phase calibrators were J1744--3116 and J1700--2610.
{The pointing calibrators  were J1745--2900, J1733--3722, J1751+0939, J1517--2422, and J1256--0547.}


The data were calibrated and reduced with the Common Astronomy Software Applications \citep[CASA;][]{mcmullin07} software package.
The calibration was performed in CASA version 5.1.2 with the calibration script provided by the East Asian ALMA Regional Center.
{Continuum emissions were subtracted from the visibility data by the task ``uvcontsub''.}
The interferometric images were created by the task ``tclean", using Briggs weighting with a robust parameter of 0.5 in CASA version 5.4.1\footnote{
CASA version 5.4.1 was used to avoid the mosaic imaging issue reported on https://almascience.nrao.edu/news/public-announcement-of-casa-imaging-issues-affecting-some-alma-products.
}.
The final image cubes of the HCN, HCO$^+$, and CS lines were generated by combining the interferometric images and TP images with the task ``feathering".
The synthesized beam sizes were as follows: $0\arcsec.93\times 0\arcsec.76$ with a position angle (PA) of $-27\arcdeg.3$ at 354.5 GHz,  $0\arcsec.93\times 0\arcsec.76$ with a PA of $-28\arcdeg.1$ at 353.6 GHz, and $0\arcsec.95\times 0\arcsec.77$ with a PA of $-28\arcdeg.3$ at 342.9 GHz.
The velocity resolution was 2 km$^{-1}$ and the rms noise levels of the HCN, HCO$^+$, and CS cubes were 7, 9, and 6 mJy beam$^{-1}$, respectively.


\begin{figure*}
\begin{center}
\includegraphics[width=16cm]{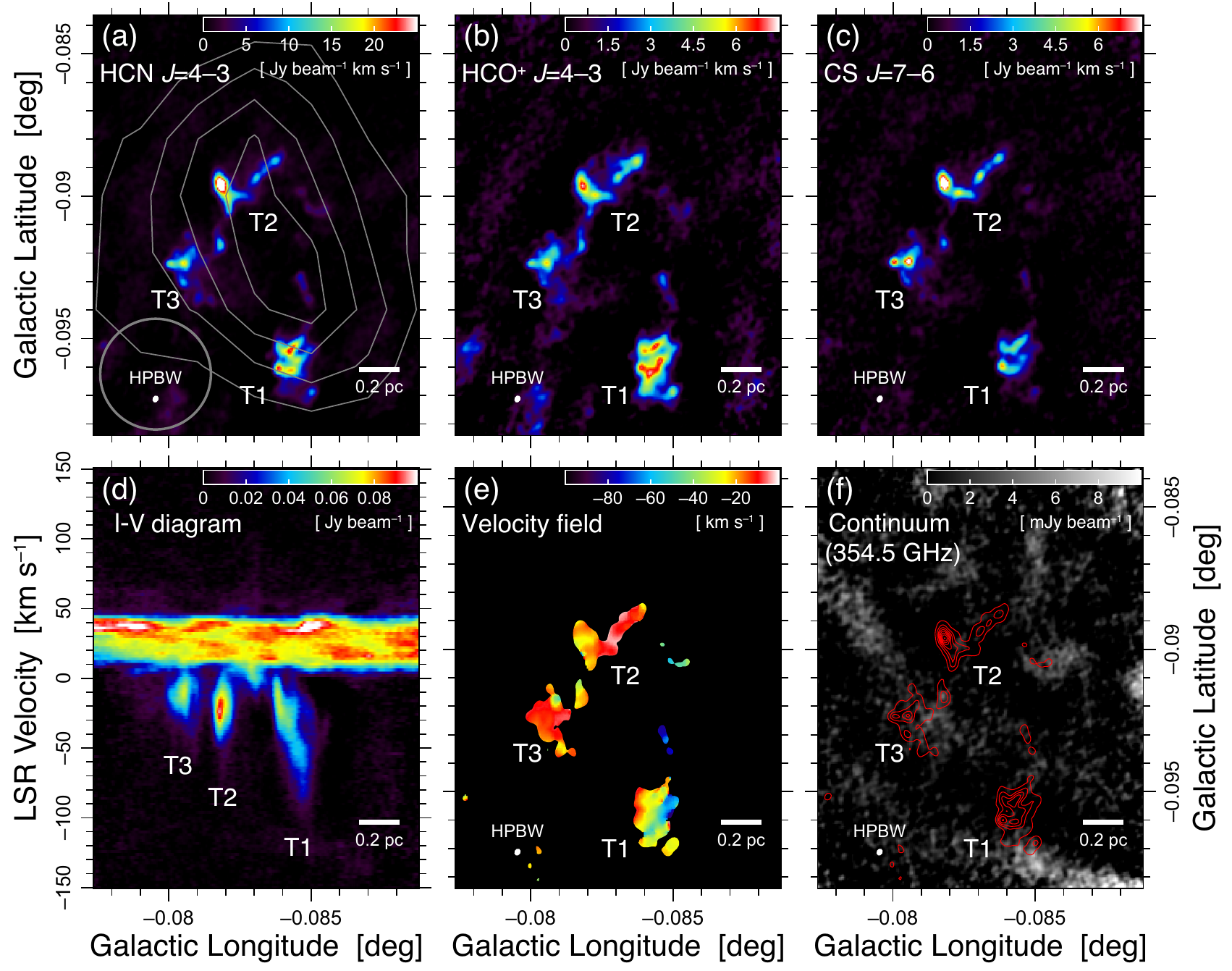}
\caption{(a)--(c) Integrated intensity maps of the HCN {\it J}=4--3, HCO$^+$ {\it J}=4--3, and CS {\it J}=7--6 lines over $V_{\rm LSR}$= $-120$ to 0 km s$^{-1}$.
The gray contours in the panel (a) show the HCN integrated intensities (10, 15, 20, and 25 K km s$^{-1}$) obtained by using the James Clerk Maxwell Telescope \citep[JCMT;][]{takekawa17b}. 
The half-power beam widths (HPBWs) of ALMA and JCMT are represented by a white filled ellipse and gray circle, respectively.
(d) Longitude--velocity ({\it l--V}) map of the HCN {\it J}=4--3 line averaged over $b = -0.\arcdeg098$ to $-0.\arcdeg088$.
(e) Velocity field (moment 1) map of the HCN line. 
(f) Continuum image at 354.5 GHz obtained with a bandwidth of 1 GHz. The rms noise level of the continuum is 2 mJy beam$^{-1}$.
}
\end{center}
\end{figure*}

\section{Results}
\subsection{Spatial and Velocity Structure}
Our ALMA observations have delineated the structure of HCN--0.085--0.094.
Figures 1(a)--(c) show the integrated intensity maps of the HCN {\it J}=4--3, HCO$^+$ {\it J}=4--3, and CS {\it J}=7--6 lines, respectively.
The integrated velocity ranges are from $-120$ to $0$ km s$^{-1}$.
HCN--0.085--0.094 is resolved into three major clumps (labeled as T1, T2, and T3), which are clearly traced in all the three lines.
The detection of the HCN, HCO$^+$, and CS lines suggests that T1, T2, and T3 consist of highly dense gas, as their critical densities are higher than $10^6$ cm$^{-3}$ \citep[e.g.,][]{shirley15}.
{Although their spatial sizes are similar ($\sim 0.2$ pc at the Galactic center distance), their morphologies are quite distinct and complicated.
The T2 clump has the highest integrated intensities in all the three maps.
Since the upper state energies of the HCN {\it J}=4--3, HCO$^+$ {\it J}=4--3, and CS {\it J}=7--6 lines are 42.5, 42.8, and 65.8 K, respectively, the CS line can trace hotter molecular gas than the HCN and HCO$^+$ lines.
The CS map presents a more clumpy appearance.
}

Figure 1(d) shows the longitude--velocity ({\it l--V}) map of the HCN line averaged over $b = -0.\arcdeg098$ to $-0.\arcdeg088$.
The emission layer distributed over $V_{\rm LSR} \simeq 10$ to 40 km s$^{-1}$ is from M--0.13--0.08 (also referred to as the 20 km s$^{-1}$ cloud), which probably lies immediately in front of the circumnuclear disk \citep[e.g.,][]{takekawa17a}.
T1 has the broadest velocity width, which reaches up to $\Delta V \sim 100$ km s$^{-1}$ in full width at zero intensity, showing a remarkably steep velocity gradient.
Figure 1(e) shows the velocity field (moment 1) map of the HCN line.
The steep velocity gradient of T1 is apparent form east to west in the velocity field.
T2 and T3 also have broad velocity widths ($\Delta V \sim 30$--50 km s$^{-1}$) for their compact sizes, but they do not represent distinct velocity gradients.

\subsection{Counterparts}
Figure 1(f) shows the radio continuum image at 354.5 GHz obtained with the ALMA observations.
The diffuse components in the continuum image probably reflect dust emission from M--0.13--0.08.
We could not identify any apparent counterparts for T1, T2, and T3 in the ALMA continuum image.

We also compared the ALMA molecular line images with the Very Large Array (VLA) 5.5 GHz continuum image \citep{zhao13, zhao16} (Figure 2(a)), the {\it Hubble Space Telescope} ({\it HST}) P$\alpha$ emission image \citep{wang10, dong11}, and the {\it Chandra} X-ray point source catalog \citep{muno09} (Figure 2(b)).
We found that the VLA radio blobs are consistently associated with the northeastern edges of T1, T2, and T3 (Figure 2(a)).
These radio blobs have been recognized as the SE Blobs \citep{zhao16}.
Sgr A$^*$ and the central cluster reside to the northeast of HCN--0.085--0.094 (green arrow in Figure 2(a)).
The positional relationship between the molecular clumps (T1--T3), radio blobs, and central cluster suggests that the surfaces of T1, T2, and T3 are ionized by ultraviolet photons produced by young stars in the central cluster.
This indicates that T1, T2 and T3 are located within the central 10 pc of our Galaxy.

The P$\alpha$ image is shown in Figure 2(b).
The bright part overlapping with T1 in the P$\alpha$ image is attributed to the incomplete continuum subtraction of a foreground star.
Due to this problem, we cannot discriminate whether T1 has a P$\alpha$ counterpart or not.
T2 and T3 have no P$\alpha$ counterparts.
Although 19 X-ray point sources were identified in the field of view, only one source overlaps with the extent of T3 (Figure 2(b)).
Their physical relation to T1--3 is ambiguous.

\begin{figure}
\begin{center}
\includegraphics[width=8.5 cm]{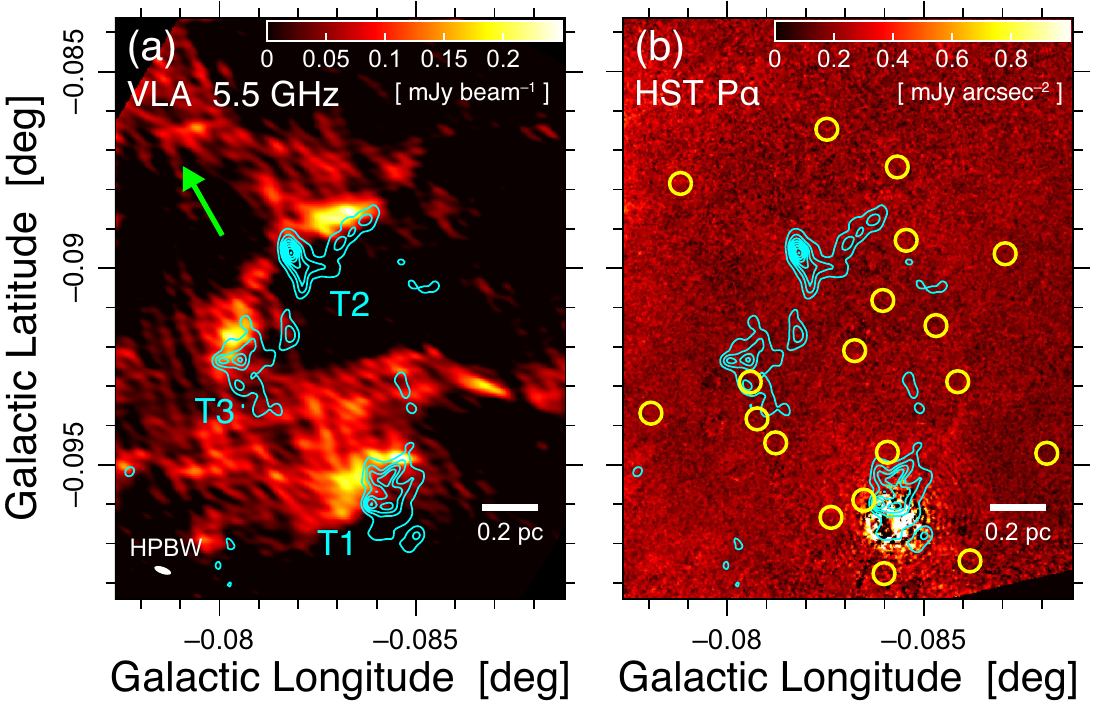}
\caption{
(a) VLA 5.5 GHz radio continuum image \citep{zhao13, zhao16}. 
The HPBW of VLA is represented by a white filled ellipse.
The green arrow indicates the direction toward Sgr A$^*$.
The cyan contours show the HCN integrated intensities (the same map as Figure 1(a)).
The contour levels are at 5 Jy beam$^{-1}$ intervals, starting with the minimum of 3 Jy beam$^{-1}$.
 (b) {\it HST} P$\alpha$ emission image \citep{wang10, dong11} . The yellow circles indicate the positions of the {\it Changra} X-ray point sources \citep{muno09}.
 The cyan contours are the same as those in panel (a).
}
\end{center}
\end{figure}

\section{Discussion}
T1 is the most prominent object in HCN--0.085--0.094 because of its broad velocity width and its clear velocity gradient.
In the following discussion, we focus on T1.

\subsection{Origin of T1}
Figures 3(a) and (b) show the CS {\it J}=7--6 integrated intensity and velocity field maps of T1, respectively.
An open ring-shaped structure (hereafter ``T1 ring'') appears, and two antenna-like structures (hereafter ``antennae''; black lines in Figure 3(a)) are associated with the northern side.
The clear velocity gradient can be seen from east to west across the cavity of the T1 ring.
The ring-like structure with the velocity gradient suggests that the molecular gas is rotating.
The gas is probably orbiting in the gravitational potential well of a massive object.
The orbital radius and rotational velocity are roughly estimated to be $\sim0.06$ pc and $\sim 30$ km s$^{-1}$, respectively.
Therefore, a mass of $\sim 10^4$ $M_\odot$ may be hidden in T1.

Alternatively, the compactness and broad velocity width may be indicative of a bipolar outflow from a massive protostar \citep[e.g.,][]{merello13}.
However, neither sub-mm dust emission nor point-like radio sources have been detected toward T1 (Figures 1(f) and 2(a)).
The molecular gas mass of T1 was estimated to be approximately 2 $M_\odot$ according to the HCN intensity under the assumption of the local thermodynamic equilibrium (LTE) with an excitation temperature of 36 K in the optically thin limit \citep{takekawa17b} and a fractional abundance of $\rm [HCN]/[H_2] = 4.8\times10^{-8}$ \citep{tanaka09, oka11}.
The kinetic energy was estimated to be $\sim 2\times 10^{46}$ erg from the mass ($M_{\rm LTE}\sim2$ $M_\odot$) and velocity dispersion ($\sigma_V \sim 20$ km s$^{-1}$). 
If T1 is a bipolar outflow with a kinetic energy of $\sim 10^{46}$ erg, the source luminosity should be brighter than $10^4$ $L_\odot$ \citep{maud15}.
On the other hand, the {\it Spitzer} 24 $\mu$m image \citep{carey09} provides a mid-infrared luminosity toward T1 of $\sim10^2$ $L_\odot$, which is much lower than the expected luminosity ($> 10^4$ $L_\odot$).
There is no sign of a young stellar object in T1.
Consequently, the protostellar outflow scenario is not plausible for the origin of the broad-velocity-width nature of HCN--0.085--0.094.

\begin{figure*}
\begin{center}
\includegraphics[width=16cm]{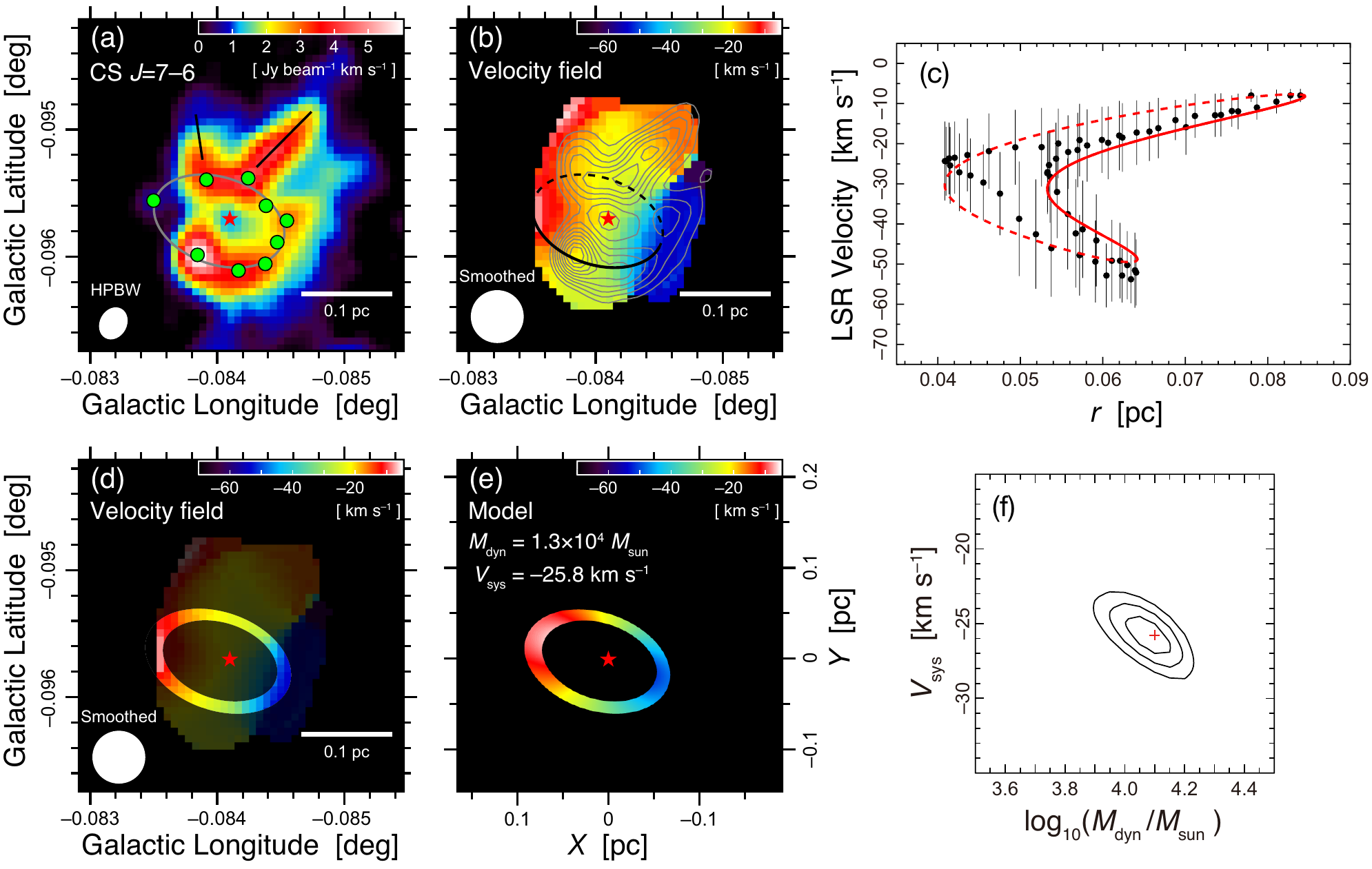}
\caption{
(a) CS {\it J}=7--6 intensity map of T1 integrated over $V_{\rm LSR}$=$-120$ to 0 km s$^{-1}$.
The white filled ellipse represents the HPBW.
The green points are the loci used for the orbital fitting for the T1 ring in the plane of the sky.
The red star indicates the dynamical center.
The gray ellipse is the modeled orbit, which is described by the orbit parameters listed in Table 1. 
The two black lines are drawn to indicate the ``antennae".
(b)  Velocity field (moment 1) map of the CS line smoothed with a $1.5\arcsec$ Gaussian kernel. 
The white filled circle indicates the full width at half maximum (FWHM) of the Gaussian kernel.
The gray contours show the CS intensities.
The dashed and solid parts of the modeled orbit (black ellipse) correspond to the dashed and solid parts of the red curve in panel (c), respectively.
(c) Position-velocity plot along the modeled orbit.
The position $r$ is a projected distance from the dynamical center.
The black points indicate the moment-1 values.
The error bars indicate the velocity dispersions (moment 2).
The red curve indicates the modeled function of the line-of-sight velocities with the best-fit values.
The dashed and solid parts of the curve correspond to the northern and southern parts of the orbit, respectively.
(d) The masked image of the velocity field of the T1 ring for comparison with the modeled velocity field.
(e) Modeled velocity field of the T1 ring with the best-fit values.
(f) Confidence level contours as a function of $M_{\rm dyn}$ and $V_{\rm sys}$. The contour levels are 1$\sigma$ (68.3\%), 2$\sigma$ (95.4\%), and 3$\sigma$ (99.7\%). The cross represents the best-fit values. 
}
\end{center}
\end{figure*}

\subsection{Keplerian Model}
The ring-like structure, velocity pattern, and absence of stellar counterparts suggest that T1 probably harbors an invisible massive object.
As is the case with the minispiral \citep{zhao09} and HCN--0.009--0.044 \citep{takekawa19a}, we performed orbital fitting to the T1 ring assuming a Keplerian motion.
We placed the dynamical center at $(l,\ b)=(-0\arcdeg.0841,\ -0\arcdeg.0957)$ and the loci used for the fitting at the locations indicated by the green points in Figure 3(a).
{We adopted a right-handed Cartesian coordinate system with the $X$-, $Y$-, and $Z$-axes parallel to the Galactic longitude, Galactic latitude, and line-of-sight direction, respectively. This coordinate definition is the same one used in \citet{takekawa19a}.}
By least-square fitting to the loci, we determined the orbital parameters: the semimajor axis ($a$), eccentricity ($e$), longitude of ascending node ($\Omega$), argument of pericenter ($\omega$), and inclination angle ($i$). 
Table 1 shows the best-fit orbital parameters.
The best-fit orbit is projected in Figures 3(a) and (b).

\begin{table}
 \caption{Parameters of the modeled Keplerian orbit}
 \centering
  \begin{tabular}{ll}
   \hline \hline 
   Parameters &    \\
   \hline
   Semi-major axis, $a$ &  $0.075 \pm 0.005$ pc  \\
   Eccentricity, $e$ & $0.19 \pm 0.06$ \\
   Longitude of ascending node, $\Omega$ & $20 \pm 10$ deg   \\
   Argument of pericenter, $\omega$ & $135 \pm 20$ deg \\
   Inclination\tablenotemark{\rm a}, $i$ &  $130 \rm \ (or\ 50) \pm 6$ deg  \\
   Pericenter distance &   $0.06$ pc  \\
   Orbital period &  $1.7\times10^4$ yr \\
   \hline
  \end{tabular}
  \tablenotetext{\rm a}{
  Orbits with $i$ and ($180\arcdeg - i$) produce the same projected orbits and line-of-sight velocities.
  If the orbital motion is clockwise, the inclination of $i > 90\arcdeg$ is chosen.}
\end{table}

After the orbital geometry was determined, we estimated the mass ($M_{\rm dyn}$) and line-of-sight velocity ($V_{\rm sys}$) of the dynamical center by fitting the observed velocities on the T1 ring with a chi-square ($\chi^2$) minimization approach.
Figure 3(c) shows the diagram of line-of-sight velocity vs. projected distance from the dynamical center of the T1 ring.
In the model fit, we used the moment-1 values of the CS image smoothed with a Gaussian kernel of $1.5\arcsec$ FWHM (Figure 3(d)) to reduce the effect of noise.
As a consequence,  we derived the best-fit values of $\log_{10}({M_{\rm dyn}}/M_\odot) = 4.10^{+0.05}_{-0.10}$ and $V_{\rm sys} = -25.75\pm1.25$ km s$^{-1}$.
Figure 3(e) shows the modeled velocity field with the best-fit parameters.
Figure 3(f) shows the confidence level contours as a function of $M_{\rm dyn}$ and $V_{\rm sys}$.
Thus, the observed kinematics of the T1 ring can be explained by a Keplerian orbit around a mass of $M_{\rm dyn}\simeq 1.3 \times 10^4$ $M_\odot$ with a line-of-sight velocity of $V_{\rm sys} \simeq -26$ km s$^{-1}$.

The antennae, which are roughly perpendicular to the T1 ring, are not considered in the Keplerian model. 
They are smoothly connected to the T1 ring in the {\it l}--{\it b}--{\it V} space with slight velocity gradients (Figure 3(b)).
The gas on the northern side of the T1 ring tends to deviate from the Keplerian orbit (see Figure 3(c)--(e)).
The antennae could be streams moving on different orbits around the invisible mass.
The velocity deviation in the northern side of the T1 ring may possibly be caused by collision with these streams.
Figure 4 shows the three-dimensional schematic view of T1.

\begin{figure}
\begin{center}
\includegraphics[width=8 cm]{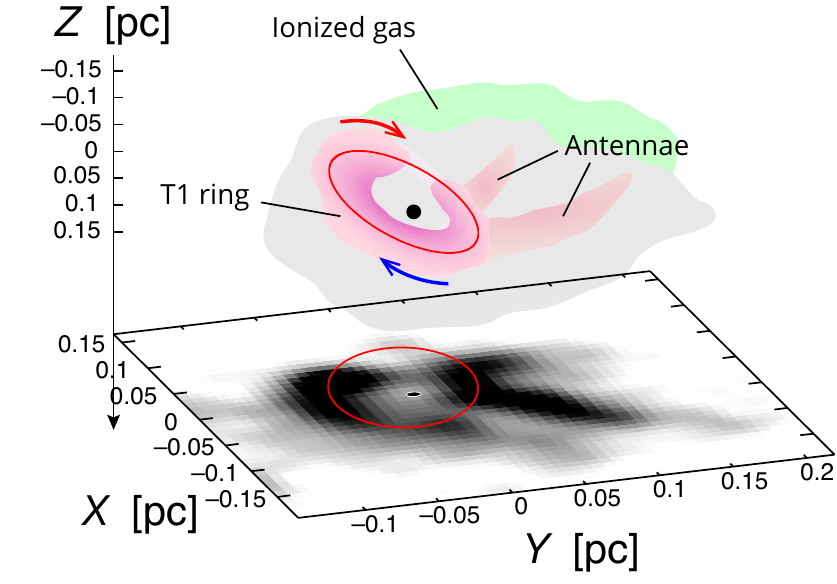}
\caption{
Three-dimensional schematic view of T1.
The red ellipse indicates the modeled Keplerian orbit with the best-fit parameters listed in Table 1.
We adopt clockwise motion ($i > 90\arcdeg$) for the orbit.
The red and blue arrows indicate the direction of the orbital motion. 
The pink clouds represent the T1 ring and antennae.
The gray and green represent the molecular gas with less column density and the ionized surface traced by the 5.5 GHz continuum, respectively (see also Figure 2(a)).
The black circle indicates the position of the IMBH.
The grayscale on the {\it X}--{\it Y} plane is the CS integrated intensity map.
}
\end{center}
\end{figure}

\subsection{Indications}
Our results indicate the presence of a dark mass of $1\times 10^4$ $M_\odot$ within a radius smaller than 0.06 pc in T1. 
The mass density of the gravitational source is much larger than $1\times 10^7$ $M_\odot$ pc$^{-3}$.
It may be an inactive IMBH because of the absence of luminous stellar objects.
IMBHs have been proposed to be formed through runaway collisions of stars in dense clusters \citep[e.g., ][]{portegieszwart02,fujii09}.
If the IMBH in T1 was born and grown in such a process, the parent cluster mass may have been $\sim 10^7$~$M_\odot$ \citep{gurkan04}, which is in the mass range of giant globular clusters or dwarf galaxies.
The possible presences of massive IMBHs have been reported in several globular clusters and dwarf galaxies \citep[e.g.,][]{noyola10, reines13, baldassare15, perera17}.
The IMBH in T1 could have been brought in by a massive cluster that migrated into the Galactic center by dynamical friction.
Many members of the parent cluster have possibly been dispersed \citep{arca-sedda18}.
Another possible origin of the IMBH is a remnant of ``seeds" for supermassive black holes in the early universe, which could be formed in the first star clusters \citep{sakurai17}.

The sensitive measurement of stellar motions in T1 may provide more support for the presence of an IMBH.
In the steep potential well, large density of stars with high velocities are expected.
Although the IMBH would currently have a low accretion rate, it could show  short-duration variabilities like Sgr A$^*$.
The short-timescale (several tens of minutes to a few hours) variabilities of Sgr A$^*$ have been detected from mm-wave to X-ray \citep[e.g.,][]{yusef-zadeh09, eckart12}.
Assuming a variability timescale of a black hole is proportional to its mass, on the analogy of the Sgr A$^*$ flashes, the variation timescale of the IMBH is expected to be a few tens of seconds at infrared and X-ray wavelengths, and a few minutes in the sub-mm.
The detection of such a very short-timescale variability from a point-like source would be a strong indication of an IMBH.
Further detailed multi-wavelength observations will enable the confirmation of the presence of an IMBH.

The ALMA observations of the HVCC HCN--0.085--0.094 led us to the discovery of a molecular clump rotating around an invisible mass, which may be another IMBH candidate.
However, there could be diffuse mass distribution around the point-like mass.
The alternative explanation for the dark mass concentration is a highly dense cluster consisting mostly of dark stellar remnants, such as neutron stars and stellar mass black holes \citep{banerjee11}.
The virial mass for the whole of HCN-0.085-0.094 was estimated at $\sim 2\times10^5$ $M_\odot$, calculated using the size parameter ($S\simeq 0.3$ pc) and velocity dispersion ($\sigma_V\simeq 20$ km s$^{-1}$).
This means that a single gravitational source with a mass of $\sim 1\times 10^4$ $M_\odot$ cannot explain all of the kinematics of this HVCC (including T2 and T3).
Detailed analysis for the other components besides T1 is left to future work.
Although the possibility of a cluster of dark stellar remnants cannot be ruled out yet, such a dense cluster could ultimately be a birthplace for an IMBH \citep{giersz15}, where gravitational waves from IMBH--stellar mass black hole mergers could be detectable through next-generation instruments such as the {\it Laser Interferometer Space Antenna} and Einstein Telescope \citep[e.g.,][]{fragione18}.

{
\section{Conclusions}
We conducted the high-resolution observations toward the HVCC HCN--0.085--0.094 in the HCN {\it J}=4--3, HCO$^+$ {\it J}=4--3, and CS {\it J}=7--6 lines with ALMA.
The inner structure of HCN--0.085--0.094 was uncovered with an angular resolution of $1''$, which corresponds to $0.04$ pc at the Galactic center distance.
The main conclusions are summarized as follows:
\begin{enumerate}
\item{
HCN--0.085--0.094 was resolved mainly into three clumps (T1--3), each of which has a broad velocity width.
The T1 clump is the most prominent in the resolved components, showing the broadest velocity width ($\Delta V \sim 100$ km s$^{-1}$ in full width at zero intensity) and very steep velocity gradient.
}
\item{
The 5.5 GHz radio blobs were found to be associated with the northeastern edges of the three clumps, suggesting that the surfaces of the clumps have been ionized by ultraviolet photons.
The positional relationship indicates that the ionizing sources are probably young stars in the central cluster.
This implies the close proximity of the three clumps to Sgr A$^*$ ($\lesssim 10$ pc).
}
\item{
The open ring-shaped structure of the T1 clump (T1 ring) and its velocity pattern suggest an orbital motion of molecular gas  around a mass of $\sim 10^4$ $M_\odot$, with a rotational radius of $\sim 0.07$ pc and a rotational velocity of $\sim 30$ km s$^{-1}$.
Because there are no bright stellar counterparts, this massive source may be an inactive IMBH, which is the fifth candidate for an IMBH in the Galactic center.
}
\end{enumerate}
}
\acknowledgments
This paper makes use of the following ALMA data: ADS/JAO.ALMA\#2017.1.01557.S.
 ALMA is a partnership of ESO (representing its member states), NSF (USA) and NINS (Japan), together with NRC (Canada), MOST and ASIAA (Taiwan), and KASI (Republic of Korea), in cooperation with the Republic of Chile. The Joint ALMA Observatory is operated by ESO, AUI/NRAO and NAOJ.
We are grateful to the ALMA staff for conducting the observations and providing qualified data.
We are also thankful to the anonymous referee for helpful comments and suggestions that improved this paper.
This study was supported by JSPS Grant-in-Aid for Early-Career Scientists Grant Number JP19K14768.
\facility{ALMA}
\software{CASA \citep{mcmullin07}, Astropy \citep{astropy13, astropy18}, Matplotlib \citep{hunter07}, Numpy \citep{vanderwalt11} }

\end{document}